# Intriguing behavior when testing the impact of quotation marks usage in Google search results

**Bogdan Vasile Ileanu**[(a) 1], **Marcel Ausloos**[(b)],
**Claudiu Herteliu**[(a)], **Marian Pompiliu Cristescu**[(c)]

(a) Department of Statistics and Econometrics,
Bucharest University of Economic Studies, Bucharest, Romania
Calea Dorobantilor 15-17, 010552 Sect. 1, Bucharest, Romania
e-mail: ileanub@yahoo.com, bogdan.ileanu@csie.ase.ro, claudiu.herteliu@gmail.com, hertz@csie.ase.ro

(b) School of Management, University of Leicester,
University Road, Leicester, LE1 7RH, UK
e-mail: ma683@le.ac.uk , marcel.ausloos@ulg.ac.be

(c) Faculty of Economic Sciences, Lucian Blaga University from Sibiu,
Calea Dumbravii nr.17, 550324, Sibiu, Romania
e-mail: mp.cristescu@gmail.come

8[th] **May, 2018**

**Abstract**
Internet research on search engine quality and validity of results demand much concern. Thus, the focus in our study has been to measure the impact of quotation marks usage on the internet search outputs in terms of google search outcomes' distributions, through Benford Law. The current paper is focused on applying a Benford Law analysis on two related types of internet searches distinguished by the usage or absence of quotation marks. Both search results values are assumed as variables. We found that the first digit of outcomes does not follow the Benford Law first digit of numbers in the case of searching text without quotation marks. Unexpectedly, the Benford Law is obeyed when quotation marks are used, even if the variability of search outcomes is considerably reduced. By studying outputs demonstrating influences of (apparently at first) "details", in using a search engine, the authors are able to further warn the users concerning the validity of such outputs.

*Funding*
*This work has been performed in the framework of COST Action IS1104 'The EU in the new economic complex geography: models, tools and policy evaluation'. Moreover, this paper is part of scientific activities in COST Action TD1210 'Analyzing the dynamics of information and knowledge landscapes (KNOWeSCAPE)' and COST Action TD1306 "New Frontiers of Peer Review (PEERE)".*

*Acknowledgements: Many thanks to Roxana Herteliu-Iftode and Gurjeet Dhesi who gave us many useful suggestions regarding the grammar and language style used in the paper.*

**Conflict of Interest**: *The authors declare that they have no conflict of interest.*

---

[1] Corresponding author

# 1. Introduction

Currently many search engines are available through internet, Yahoo (https://ssearch.yahoo.com/), Bing (www.bing.com), Ask (www.ask.com), etc. With more than 158 million different visitors (Nielsen, 2014), Google (www.google.com ) is the recognized leader of the world for internet searchers, being able to "produce" outcomes for any kind of person and desire. This is why Google became 2nd World's Most Valuable Brand so rapidly (Forbes, 2016). Many quantitative studies and methods to measure the quality of results concluded that the Google search engine is the most precise one (Li and Shang, 2002; Sabha and Sumeer, 2016).

Google search can return information on a huge amount of "searching products": questions arise about the outputs reliability. Any reader of this paper is aware that misprints may inadvertently occur in writing the subject line. Sometimes, one copies and pastes keywords. Thus, it can be asked how special style signs affect, refine or dilute the quantity and quality of wished results. Google support helps the users how to use the "search operators" and informs them about the difference in usage but its help is weak in numbers.

One important operator is the quotation marks: for words or phrases between quotes, the results only include pages which contain words or group of words identically spelled and in the same order with those inside the quotes. Even if the searching outcomes were found to be influenced by engine's geographical customization and being dependent on different searching markets determined by e.g. location, language and other characteristics (Wilkinson and Thelwall, 2013), the impact of quotation marks has not been measured, even though it provides qualitative and quantitative findings, suggesting supplementary thoughts. In fact, quotation marks are recommended to be used in order to check the reliability of an internet information (RTBF, 2016). Furthermore, any reader also knows that, even though the search question should be very precise, one has often to reduce the number of keywords in order to provide qualitative and quantitative findings on an acceptable finite size scale (Bo and Chunming, 2013).

Beside the fact that social life is today strongly connected with Google search outcome, the academic research studies are using more and more results of search engines, therefore some analysis regarding these engines outputs becomes necessary.

Being aware of caveats, pointed already by many authors, that search engines are manipulated by providers without explicit information to the questioners, or that searching results are not always reliable due to different reasons, we study here especially the Google search results, in certain conditions of interrogations, delimited by the usage of quotation marks, using the Benford Law (1938) (BL).

We organized the paper in three major section: literature review, methodology & results and discussion and conclusions.

# 2. Literature review

## 2.1. Internet searching and relevant quantitative results

Being as part of our life, internet searching and quality of the results is a frequent topic also in the scientific studies (Paulheim, 2014; Evans, 2007, etc.). From the multitude of these studies only a few are related to our question "on outcome reliability". We are briefly describing the most relevant ones among them in the next paragraphs.

The outcome of search engines has been widely approached. A pioneering paper is due to Huberman *et al.* (1998) who using multiple queries and/or simulations found that many results of the activity on the internet respect known natural laws such as inverse Gaussian or Zipf's laws (Zipf, 1949). For example, the impact of Chinese characters on web-search queries has been found to lead to power laws (Chau *et al.*, 2009). Other authors compared different search engines and their quality of search returns (Wang *et al.*, 2012). Wouters *et al.* (2004) remarked that the results of search are tightly dependent of time course: old information being less accurate returned or missing. In Pimenov and Ilyin (2010) special characters are used and query designation in the case of complex queries, then search results accuracy is discussed. Pimenov (2008) and focus on queries restrictions with different types of keywords, e.g. headings, meanings, usage of words and revealed search engines weaknesses, like file multiplication or results instability.

Depken (2008) analysed the distribution of the most popular 500 blogs by number of incoming links using Benford Law and blogs ranking using Zipf's methodology. Both analyses reveal that such theoretical laws are not holding in those cases. Lansey and Bukiet (2009) concluded that Google outcome searches, made by different types of word combinations, follow different types of power laws (Zipf, Heap's Law); moreover, the searching outputs of random numbers follow a Benford Law.

## 2.2. Benford Law and its applications

Newcomb, observing the difference in "cleanliness" of pages with different logarithm mantissa, concluded that the occurrence probability of 1, 2, ..., 9 as the first digit of a number decreases from about 0.3 in the case of 1 to less than 0.05 in the case of 9 (Newcomb, 1881). Almost 60 years later, Benford empirically deduced the thereafter called "laws of anomalous numbers", the Benford Laws (BL) (Benford, 1938, Mir and Ausloos, 2017).

According to BL the occurrence of the digit v={0,1,2,3,4…,9} as the k-th digit, $d_k$ of a number, should appear with a probability:

p($d_k$= v) = $\sum_{k=10^{d_k-2}}^{10^{d_k-1}-1} log_{10}\left(1 + \frac{1}{10k+v}\right)$. Then we have the following particular distribution

$$\begin{pmatrix} d_1 \\ p(d_1) \end{pmatrix} = \begin{pmatrix} 1 & 2 & \cdots & 9 \\ 0.301 & 0.176 & \cdots & 0.044 \end{pmatrix}$$

which is widely known as First Digit Benford Law (BL1). The appearance probability of v={0,1,2,3,…,9} as the second digit or as the third digit can be derived theoretically, leading to so called BL2 and BL3 theoretical distributions. They tend to become more uniformly distributed as the position (k) of the digit increases.

On one hand, the BL usefulness came to prominence following Nigrini's observation about its frequent emergence in financial data (Nigrini, 1996a; Nigrini, 2015, Ausloos *et al.*, 2016, Ausloos *et al.*, 2017b, Riccioni and Cerqueti, 2018, etc.). Thereafter, the application of BL, as a test of occurrence has been used in many different fields, from: geography-geology (Sambridge, *et al.* 2010; Nigrini and Miller, 2007, Ausloos *et al.*, 2017a), astronomy (Fox and Hill, 2014), finance (Mir, 2016; Slijepčević and Blašković, 2013; Clippe and Ausloos, 2012), audit (Nigrini, 1996a; Nigrini, 1996b), scientometrics (Campanario and Coslado, 2011), religion (Ausloos *et al*., 2015; Mir 2012), voting fraud (Mebane, 2004; Roukema, 2014), informatics (Arshadi and Jahangir, 2014), to mafia effects on income taxes (Mir *et al*., 2014). The tendency is to verify the empirical distribution of different sets of values assumed to be non-manipulated with respect to a theoretical distribution given by Benford Law. Non-obedience to BL induces suspicion on the data set, suggesting anomalies or even fraud. Possible explanations are given each time. According to Nigrini and Mittermaier (1997), BL should not (or does not!) apply when human thought (or acting constraint) is involved.

On the other hand, for explaining BL1, many examples, theories, and research findings have been presented (Pinkham, 1961; Wouk, 1961; Beebe, 2016; Kossovsky, 2014). An important chapter on Benford Law research is made by mathematical research completed many times with real examples, indeed. Some fresh and debated subjects connected with our study are those on the "conditions" in which variables or data sets (have to) obey or not Benford Law. We recall here only those with major importance: Pinkham (1961) gives the principle of invariances (in scale and base) retaining from the property related to constant multiplication or unit transformation. For example, if a set of values in USD, conformant to Benford Law, are converted into some 2other currency, are still found to be Benford conformant. Berger and Hill (2011) give several distributions which are conformant to Benford Law, while Lemis, *et al.* (2000) were testing conformity of survival distribution. In the last-mentioned study for example the exponential, Weibull or Log-logistic with certain parameters are found to be BL conformant. Some important theorems and properties about Benford Law can be found also in Gauvrit and Delahaye (2009) and Berger and Hill (2011). With simulated data Forman (2010) studied the conformity of BL for exponential, half-normal, normal, etc. distribution and the ratio of two variables coming from those distributions. Gauvrit and Delahaye (2009) and Clippe and Ausloos (2012) applied different functions such as $\log(\log(u))$, $\log(\sqrt{u})$, $u*\log(u)$, etc on the initial set of real or simulated data, *u*, and then applied Benford Analysis before and after the transformation. However, many observations and deductions are puzzling (Judge and Schechter 2009).

## 3. Data Analysis

### 3.1. Methodology

For outlining the methodology and subsequent logics of our investigation, let us refocus on previous papers analysing World Wide Web data with a BL spirit. Three reports are pertinent: (a) Lansey and Bukiet (2009) and (b) Depken (2008). Thus, we offer a fourth study, an appropriate investigation with a well framed question. We decided to look for the impact of Google engine results due to the impact of quotation marks (or not) in keywords, - a precisely well-defined alternative indeed. The "keywords" have been chosen to be the names of head of states, in order to maintain the search to a reasonable size. Thus, we have recorded the number of times a head of state name is found investigating two types of search, called method A1 and method A2, without or with quotation marks respectively, in order to reveal some search bias.

The A1 method starts from the name of head of states, listed on: http://en.wikipedia.org/wiki/List_of_current_heads_of_state_and_government, with rare exceptions: we avoided the "leaders" of territories not widely recognized in the world such as: Abkhazia, Ossetia, Kosovo, ... and those of Canada, Australia or New Zeeland, ... since they are (still) under Queen Elizabeth II authority; 2 countries where voluntarily excluded due to unclear or some controversy about the leadership. Here were the case of Presidency of Bosnia and Herzegovina where there was a three-member "consensus" body which collectively serves as the "head of state" and President of Yemen Republic was also avoided since this function was at the moment of analysis rather disputed.

Other three exceptions were made because the mandate ended or began around the starting period of analysis. Here are the cases for Poland, where we searched for Bronisław Komorowski, instead of Andrzej Duda, the case of Greece, where we searched for Karolos Papoulias instead of Prokopis Pavlopoulos and the case of Croatia were we searched for Ivo Josipovici instead of Kolinda Grabar-Kitarović. We preferred to have longer periods of mandate to avoid some outliers caused when a person is a new entry.

The English language was used in general in spelling the name of presidents and countries. Exceptions procedures were made for Macedonia, Spain and Portugal for which the "local" spelling was used.

Each name of the head of state was typed in the Google search engine as a keyword: name surname president of *country name;* e.g.: Barack Obama president of United States. If the keyword was misspelled, the Google search self-correction was accepted. If the head of state is a king (e.g. Sweden), emperor (e.g. Japan), emir (e.g. Kuwait), … the word "president" was replaced with the adequate title.

In the case of A2, each keyword used in the A1 method was introduced between quotation marks, e.g., Ram Baran Yadav president of Nepal (form in A1 queries) was introduced as "Ram Baran Yadav president of Nepal" in (A2) query.

The Google search engine returned a much higher value of results in (A1) than in (A2): indeed, (A1) returns the sum of results for each word. In (A2), the quotation marks constraints strictly impose to find the exact group of words; these results tend to be more under human control.

All these search activities were performed in less than 12 hours, in 2016, in order to minimize the bias caused by possible changes of the Google results naturally determined by time course and Web documents reorganization by the output provider in their databases (Wouters *et al.*, 2004).

Finally, just to crosscheck the validity of information collected from Wikipedia, on the possible head of state, we selected a random sample of ten less well known countries and head of states, according to our knowledge: Nauru, Gambia, Burundi, Kiribati, Guyana, Comoros, Bulgaria, Ghana, Haiti and Trinidad and Tobago. The accuracy of the name of the mentioned list of heads of state, using the official presidency homepages was crosschecked. No anomaly was found.

Nevertheless, the English language is expected to affect the frequency of occurrence results in a search method. For example, countries where the national language is not English and were not ever territories of UK or USA dominance had small chances to post searchable message about the presidency in the English language. We point to examples like Congo, Brazil, Comoros and Angola. It is commonly accepted indeed that the nationalism of French and Germans are downgrading the English language in search engines; for this reason, low values of the number of these heads of state can be expected in counting search outputs.

We must recall that the precision of the data is time dependent, as expected in such studies. Moreover, search engines are known to guess (and to remember) the answer the researcher wishes. Therefore, some data bias is integrated beyond the researcher will into the outputs. Thus, even if the engine's outcome number of findings is not quite exact, - in fact, the displayed results are preceded by the word "about" (!), as the reader surely has noticed, the bias does not appear to be sufficiently significant as to impeach the analysis and its discussion. Based on various tests, we have estimated that the deviation appears to be of the order of 1 000 on 1 000 000 "results", whence allowing anyone to consider that the "statistical error" ($10^{-3}$) is obviously not large enough to influence the value of the first digit in the number of items found on such a Google search.

### 3.2. Descriptive analysis

The numerical range of search outcomes for a given head of state name, is presented in figure 1, - revealing that a large number of heads of states (49 among 169 investigated) return low values (less than 100 thousands). The 3 lower results are for the president of Macedonia, Laos, and Samoa. In this range below 100 000, one finds Bolivia, Equatorial Guinea, Guinea Bissau, Mauritius Islands, Peru, Sao Tome, Nauru, and surprisingly, for us, the Bulgaria president, with a 2450 "count".

On the right part of the distribution in contrast, we can observe the presence of a significant number (19) of head of states for which the Google search is returning more than 600 000 counts. The top three results somewhat could be expected: President of USA, (more than 93 billion results), President of Russia and Xi Jinping, President of PR China, but also some other "popular" leaders like: President Bashar Al Assad, Syria, King Salman from Saudi Arabia, Jacob Zuma from South Africa have also top values.

In the case of A2 method, the search results dramatically drop. For example, the maximum number of results is less than 400 000 in contrast to the first analysis in which the maximum reached the (high) value of 93 500 000; see table 1.

The presence of quotation marks in A2, determines a change in the ranking order of the head of states. The top 3 are now represented by the President of Palestine Mahmoud Abbas, Hassan Rouhani, president of Iran, and Mohammed VI, King of Morocco.

The bottom results, in A2, unexpectedly remain quite stable and given like in the A1 methods case by: Samoa, Laos, but with RD Congo head of state, instead of the Macedonia president.

**Table 1.** Descriptive statistics regarding Google search outcomes along both methods of search discussed in the text (rounded up values)

| Indicator | First analysis (method A1) (simple text) | Second analysis (method A2) (simple text between " ") |
|---|---|---|
| Mean | 1 119 695 | 14 734 |
| Standard Error | 577 788 | 3 174 |
| Median | 237 000 | 3 710 |
| Mode | 148 000 | 2 890 |
| Standard Deviation | 7 511 238 | 41 268 |
| Kurtosis (K) | 139 | 32 |
| Skewness(S) | 11 | 5 |
| Range | 93 499 777 | 330 000 |
| Minimum | 223 | 0 |
| Maximum | 93 500 000 | 330 000 |
| Number of searches | 169 | 169 |

The major "…" impact is given when both ranges are compared. In A1, the number of outcomes are lying on a huge interval measuring 93,499,777 units, almost 300 times larger than the interval size when quotation marks, " ", are used. Due to this variation all main statistical indicators, mean, median, mode reach very different values. In A2, the minimum number of searches outcomes is 0 corresponding to Choummaly Sayasone Shiavong from Laos which seems to be "invisible" on www.google.ro, but this particular result it is not a surprise since when we used the A1 method, he was also the least "visible" with only 223 Google results.

The impact of quotation marks is also significant when homogeneity is studied. In the first analysis (A1), the variation coefficient was 6.70 (670%) showing a very high heterogeneity. For the second analysis type (A2), the heterogeneity is still very high but the variation coefficient was only 280%, showing a significant reduction. The shape parameters K and S values indicate that, in both cases, the empirical distributions are very far from the Gaussian distribution and due to a very large variance of responses in both methods of search combined with the power law distribution visible in Figure 1 and Figure 2, the results are theoretically qualifying for a Benford Law analysis.

### 3.3. The quotation marks impact and Benford Law Analysis

The raw data are given in the Annex. The distribution of primary data is examined for the first and the second digits: Benford Law conformity plots are reported in the Figure 1 in the case of A1 and in the Figure 2 for the second type of analysis (A2).

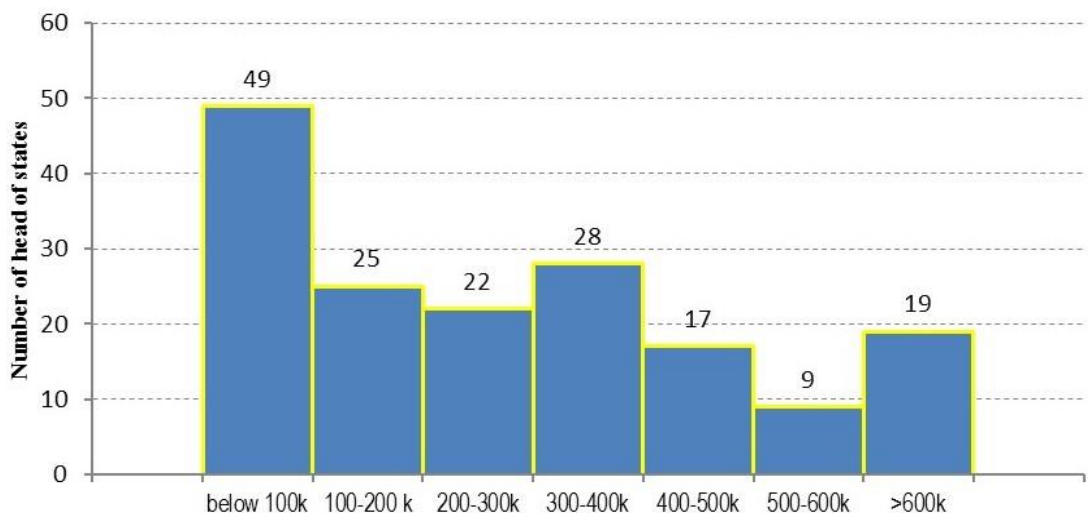

**Figure 1.** Number of Google search outcomes without quotation marks

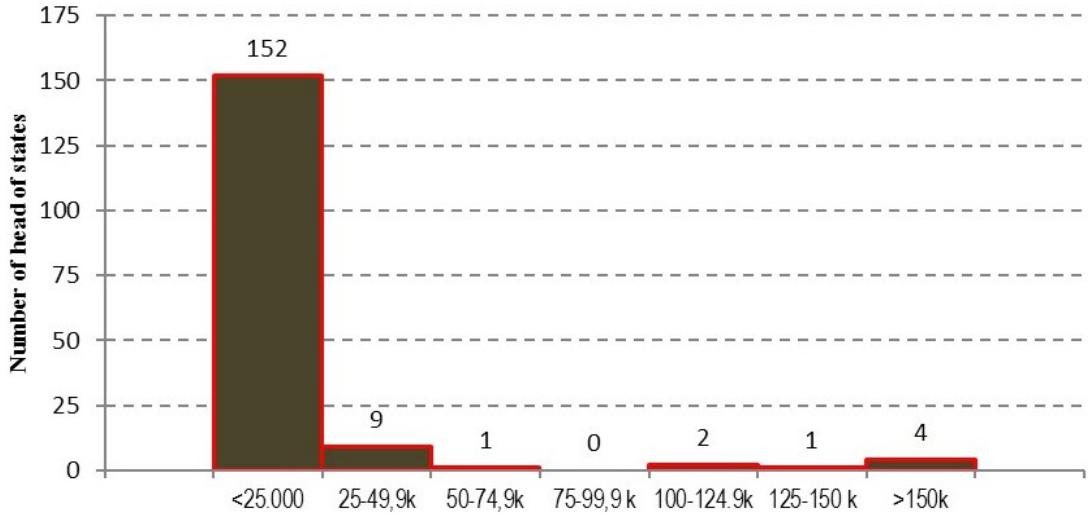
**Figure 2.** Number of Google search outcomes with quotation marks

The Benford Law analysis described in the 2.2. sub-section above and performed on the data pictured in Figures 1 and 2, gives the possibility to test the following couple of hypotheses in the case of both search methods A1 and A2, and for digits in two different positions, (first and second place, in this case), respectively:

H$_0$ *Distribution of the first digit/second of each searching outcome conforms to Benford Law*

H$_1$ *Distribution of the first digit/second of each searching outcome is not conforming to Benford Law, thus human interaction has a great impact on Google search outputs when the most powerful people, the head of world states, serve as keywords.*

To test whether the null hypothesis holds or not, the chi-square, $\chi_{c1}^2 = \sum_{i=1}^{9} \frac{(n_{oi1}-n_{ei1})^2}{n_{ei1}}$, where $n_{oi1}$ is the observed absolute frequency of $i$ digit occurred as first digit and $n_{ei1}$ is the expected absolute frequency of the same digit $i$ to appear on the first position, was applied.

    (i)    the case of searching method A1 (no quotation mark used)

In this case when the empirical value of chi-square is computed, for the first digit analysis, the achieved value is 21.92. At the 0.05 level of significance, for 9-1=8 degrees of freedom, the expected chi-square is 15.51. One can conclude that there is a statistically significant difference between the observed and theoretical distributions. A visual analysis of the differences between the two distributions, represented in the Figure 3, is revealing that only the frequency of 1 and 3 occurrences as first digit are enough to perturb the BL1 conformity, the other points are well distributed and do not contribute much to the chi-square value.

In the case of second digit analysis, even if the 0 and 1 digits' occurrence is quite different as compared to the expected BL2 the" … " impact is not strong enough to carry the empirical value of the $\chi_{c2}^2 = \sum_{i=0}^{9} \frac{(n_{oi2}-n_{ei2})^2}{n_{ei2}}$ over the 0.05 level of significance. Since the computed chi-square is 12.32 and the critical value at 5% risk for 9 degrees of freedom is 16.92, we assume BL conformity in this case.

(ii)    the case of search method A2 (the presence of quotation marks)

On the second analysis, when the outcome of the Google search is analysed for a more accurate phrase related to a head of state, the computed chi-square has a value equal to 6.24

The small differences between empirical and theoretical results, pictured in the figure 4 are statistically non-significant; thus, we accept, at 0.05 level, that there is no significant difference between the empirical distributions and the theoretical one defined by BL1 in the case of first digit.

The *second digit* empirical distribution presents also small deviations from the theoretical (or expected value), especially in the case of digit 1 and 5. But since the computed value of the chi-square, here being 10.21, is below the theoretical one, we do not reject the null hypothesis either in this case. Thus the empirical distribution follows the BL2, sustaining the hypothesis of non-anomalous search results even if quotation marks are used.

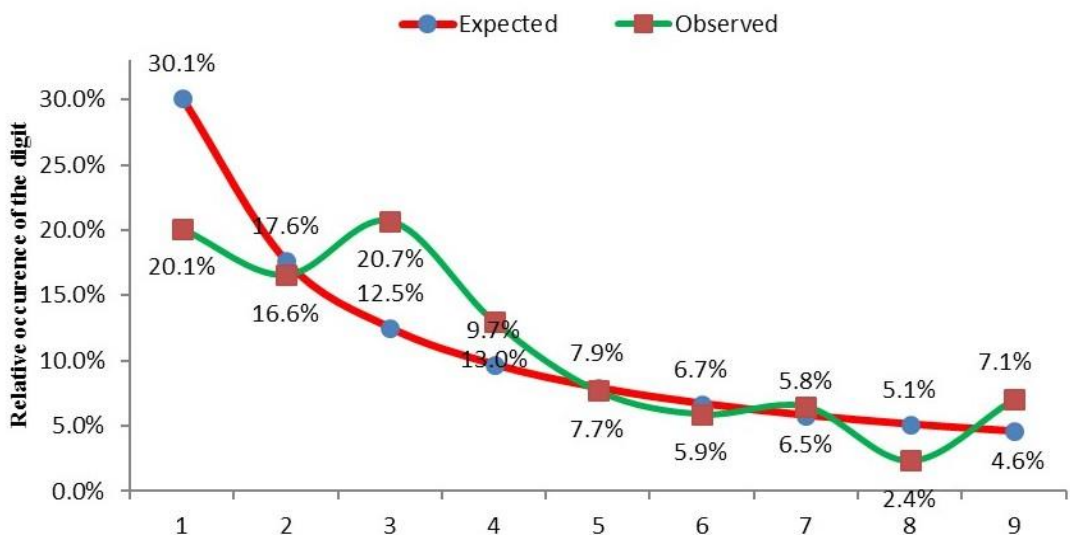

**Figure 3.** Expected versus observed relative frequencies for First Digit

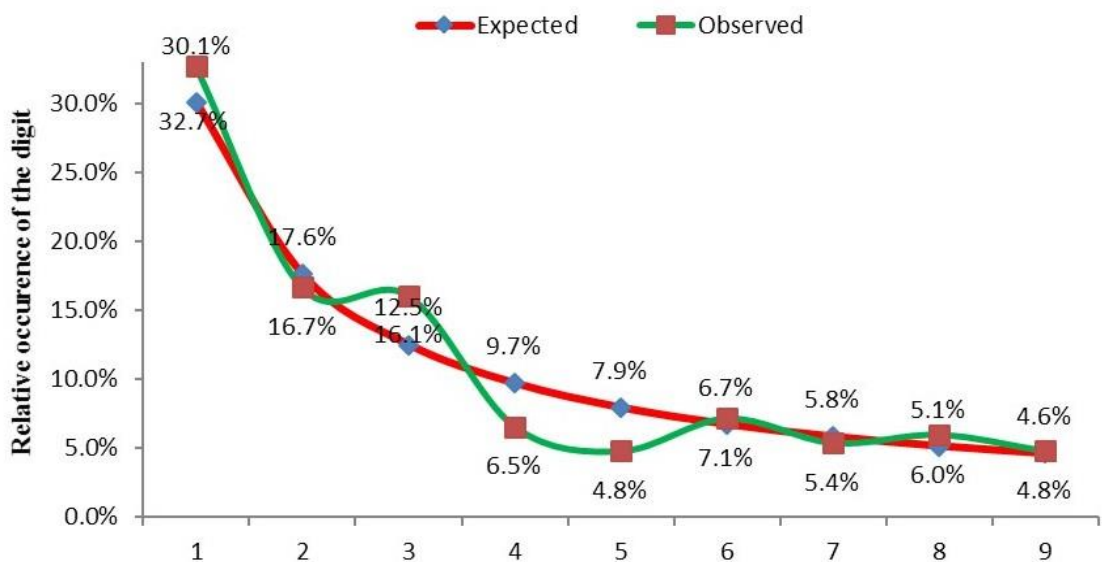

**Figure 4.** Expected versus observed relative frequencies for Second Digit

The reader can conclude that the main anomalous result achieved is met when BL1 is tested. In Figure 3, it is obviously seen that the problem arises at digit 1 (under occurred) and digits 3 and 9 over occurred. Theoretically it is possible that by eliminating the top x% of the top/bottom values to solve the problem from digits 3 and 9 and as a result to reduce the overall computed chi-square.

Thus considering that the anomalous result is caused by the presence of outliers we first delete the top 5% cases meaning about nine outcomes. The top nine values achieved when the method A1 is used have 9 as first digit in 5 cases, 1 in two cases, and digit 2 and 3 in one case each. Unfortunately, the exclusion of the top 5% of the values is not a solution in this case, mainly because the discrepancies from digit 1 and 3 are not corrected. However, ignoring the top 5% values creates a small advantage reducing the discrepancy at digit 9. However, this outlier removal is not enough sufficient to lower the computed chi-square below the critical value.

Another possibility is to ignore the bottom 5% of the values. The smallest 9 values in the case of A1 method of search are: 15900, 15100, 14100, 13500, 4980, 2450, 2110, 1540, 223. The presented values are dominated by the values which have first digit 1; the ignorance of these does not add the expected impact on the computed values. There is no significant improvement on BL1 test results increased significance if 1% or 10% of cases are deleted, mainly because of the digit 1 and 3 major discrepancy which are not reduced when values are deleted.

## 4. Discussion and conclusions

### 4.1. Discussion
In summary, recall that we consider two related types of information for the Google search engine: in brief, the name of the head of state either spelled without or with quotation marks, serves as the keywords in analysing search bias. There were 169 states "of interest". The frequency outcomes are shown to be quite different. The results were tested in Benford Law sense to see how "natural" or contrary, anomalous, are the results spread over the countries.

We find that the first digit of the outcomes *breaks down* the Benford Law in the case of searching text without quotation marks, but surprisingly, when quotation marks are used, in A2, the first digit distribution *obeys* Benford Law.

We emphasize that this result is contrary to the usual expectations since the Google searches in A2 are the results of a stronger human interaction than in A1 case, which, as discussed in the literature, is thought to impair Benford law validity. Moreover, the second digit still better fulfils Benford Law in both cases (methods A1 and A2).

The analysis was enhanced here borrowing methods from other fields and applied to internet research outcomes.

It might be interesting to imagine reasons for the above findings. In the first part of the Google outcome analysis, the search engine gives results (on names) much depending on the current political situation (e.g. Syria crisis, Russia crisis, US President position regarding all world context) on the globe, - since each of these actors is the head of state of the most important countries in the world.

Other factors which influence the number of Google search outcomes can be found in (i) the large population of the president's country, (ii) the complexity of spelling the president's name, (iii) the complexity of spelling the country/state's name. Indeed, there is likely a lower probability to enter on the Google search box the names such as "Mohamed Ould Abdel Aziz" or "Hery Rajaonarimampianina" instead of "Macky Sall", King Salman or "Robert Mugabe". Also, the notoriety of a head of a state is likely dependent of the country government form and implicitly depends on the head of state attribution according to the type of political organization. In fact, a rapid search shows that Angela Merkel as Chancellor of Germany will be found much more frequently by a Google search engine: 93000 results as compared with 12200 results, for the president of the same country, "only" because of her attribution and her notoriety. The length of mandate is also likely a very important determinant of search outcomes. For example, the president of Romania, Klaus Johannis who was elected in November 2014 or presidents of Mozambique and Zambia who were elected in January 2015 were expected to give fewer results than the leaders of which mandate are close to the finish or even finished at the moment of the analysis or at the moment of writing the paper (eg. Bronisław Komorowski from Poland, Karolos Papoulias from Greece, etc.).

## 4.2. Conclusions

We have attempted to find out the role of "details" on search output pertinent to information science, through a non-trivial analysis based on Benford law. Many applications relating Benford law were listed in the first section of this paper. In such a large amount of papers, it is easy to expect also the presence of debates on when Benford Law should be applied and what types of distributions obey this "natural law". Basically many researchers recommend not to be testing the Benford Law on distributions known to be a result of delivered human interaction or human constraints such as quantities sold, telephoning numbers, etc. A frequent application of Benford law tends to be directed towards detecting fraud or anomalies/errors. Thus, at the beginning of this study we thought that (1) surfing on the internet is an activity where human interaction is a main factor decisive for the results, and (2) resulting search outputs regarding the head of states, as the most important or powerful people in the state and in the world, might be affected by wanted or unwanted human manipulation. In such cases, we rather expected that Benford Law conformity was going to be rejected in both cases or at least in the second case, when quotation marks are used.

Beside many other caveats on the reliability of search engines, the search outcomes dramatically differ according to a detail in the keywords-the presence of quotation marks. It is true that the findings are for a specific search, a given president and made from a specific computer- authors' but this study shows that the "popularity" of the presidents on the internet is dramatically affected by quotation marks. We indicated that reasons of the variability outcomes presented above, the president' countries' language(s) must be also taken into account. However, the difference in outcomes given by the two methods of search is so huge that other factors than the presence of quotation marks will have a small impact.

The frequency of the first and second digit of search outcome numbers which is found to behave anomalously and unexpectedly raises a question about the quality of Google searching outcomes and their impact on the current socio-political life.

Our analysis shows that Benford Law obedience is rejected in the first case and accepted in the second case pointing out to a possible controversy, - or data fraud (Mir *et al*. 2014), - in our view, a Google anomalous behaviour of the results. We have to acknowledge the existence of unknown cofounders which bothers enough the results given by Google search before and after the quotation marks usage.